
%
%
\input texb
\input epsf
\magnification=1200\overfullrule=0pt\baselineskip=12pt
\vsize=22truecm \hsize=15truecm \overfullrule=0pt\pageno=0

\font\titlefont=cmbx12 scaled \magstep1
\font\sectnfont=cmbx10  scaled \magstep1
\font\subsectnfont=cmbx8  scaled \magstep1
\font\eightrm=cmr8
\long\def\fussnote#1#2{{\baselineskip=9pt
     \setbox\strutbox=\hbox{\vrule height 7pt depth 2pt width 0pt}%
     \eightrm
\footnote{#1}{#2}}}
\def\mname{\ifcase\month\or January \or February \or March \or April
           \or May \or June \or July \or August \or September
           \or October \or November \or December \fi}
\def\date{\hbox{\strut\mname \number\year}}
\newcount\REFERENCENUMBER\REFERENCENUMBER=0
\def\REF#1{\expandafter\ifx\csname RF#1\endcsname\relax
               \global\advance\REFERENCENUMBER by 1
               \expandafter\xdef\csname RF#1\endcsname
                         {\the\REFERENCENUMBER}\fi}
\def\reftag#1{\expandafter\ifx\csname RF#1\endcsname\relax
               \global\advance\REFERENCENUMBER by 1
               \expandafter\xdef\csname RF#1\endcsname
                      {\the\REFERENCENUMBER}\fi
             \csname RF#1\endcsname\relax}

%
%
\newcount\SECTIONNUMBER\SECTIONNUMBER=0
\newcount\SUBSECTIONNUMBER\SUBSECTIONNUMBER=0
\def\section#1{\global\advance\SECTIONNUMBER by 1\SUBSECTIONNUMBER=0
      \bigskip\goodbreak\line{{\sectnfont \the\SECTIONNUMBER.\
#1}\hfil}
      \bigskip}
\def\subsection#1{\global\advance\SUBSECTIONNUMBER by 1
      \bigskip\goodbreak\line{{\subsectnfont
         \the\SECTIONNUMBER.\the\SUBSECTIONNUMBER.\ #1}\hfil}
      \smallskip}
%
\def\lsim{\raise0.3ex\hbox{$<$\kern-0.75em\raise-1.1ex\hbox{$\sim$}}}
\def\gsim{\raise0.3ex\hbox{$>$\kern-0.75em\raise-1.1ex\hbox{$\sim$}}}
\def\banner{\hfill\hbox{\vbox{\offinterlineskip
                              \binum\uctnum}}\relax}
\def\manner{\hbox{\vbox{\offinterlineskip
                        \uctnum\binum\date}}\hfill\relax}
\footline={\ifnum\pageno=0\manner\else\hfil\number\pageno\hfil\fi}

\def\uctnum{\hbox{UCT-TP 199/93 \strut}}
\def\banner{\hfill\hbox{\vbox{\offinterlineskip
                              \uctnum}}\relax}
\def\manner{\hbox{\vbox{\offinterlineskip  }}\hfill\relax}
\footline={\ifnum\pageno=0\manner\else\hfil\number\pageno\hfil\fi}
\def\UCT{Department of Physics, University of Cape Town, Rondebosch
7700,
South Africa}
\def\BI{Fakult\"at f\"ur Physik, Universit\"at Bielefeld,
        W-4800 Bielefeld 1, Germany}
\def\Wroclaw{Institute of  Theoretical Physics, University of
Wroclaw, Poland}
\def\Kur{Kurchatov Institute of Atomic Energy, 123182 Moscow, Russia}
{\vsize=21 truecm\banner\bigskip\baselineskip=10pt
\begingroup\titlefont\obeylines
\hfil Influence of the Landau-Pomeranchuk Effect\hfil\smallskip
\hfil on Lepton-Pair Production$^*$.\hfil
\endgroup
\bigskip
\centerline{
J.~Cleymans$^{1}$,~V.~V.~Goloviznin$^{2,3}$ and K.~Redlich$^{2,4}$}
}
\bigskip
\noindent {$^1$}\UCT\hfil\par
\noindent {$^2$}\BI\hfil\par
\noindent {$^3$}\Kur\hfil\par
\noindent {$^4$}\Wroclaw \hfil\par
\bigskip\centerline{\bf Abstract}\medskip
{\eightrm\baselineskip=8pt
 An estimate is made of the Landau-Pomeranchuk effect on the
production of dileptons in a hadronic gas and in a quark-gluon
plasma.
For low mass dilepton pairs this effect
reduces the  production rate for bremsstrahlung by an order of
magnitude.
For  high invariant masses its influence
is negligible. This behaviour is of importance for the
theoretical
analysis of low mass dilepton pairs produced in relativistic heavy ion
collisions.}
\bigskip\bigskip
\section{Introduction}
The prime objective of ultra-relativistic heavy ion
collisions is to find physical observables sensitive to the phase
transition from hadronic matter to a quark-gluon plasma. It has been
suggested that
dilepton pairs would be one of the best signals for this
[\reftag{vesa}].
In order to describe  them one needs to estimate  quantitatively
dilepton production in a hot hadronic medium
and in a quark-gluon plasma.

It is well known [\reftag{crs}] that low invariant
mass dilepton pairs mainly come from
\bigskip
\hrule\medskip
\noindent $^*${{\sl Talk presented at the
``Workshop on Pre-Equilibrium Parton Dynamics in Heavy Ion
Collisions.''

\noindent LBL, Berkeley, August 23 - September 3, 1993}}
\eject
Dalitz decays and from
bremsstrahlung. Figure (1) shows the relative importance of various
dilepton sources, in particular,
it can be seen that bremsstrahlung from pions is one of the most
copious
sources.\par
\bigskip
%
Figure 1 : Contributions of different processes to low mass
dielectrons.
\bigskip
It was however pointed out subsequently that
medium effects
could substantially modify this rate. The present work was motivated
by the fact that bremsstrahlung in a medium is suppressed by the
Landau-Pomeranchuk [\reftag{landau},\reftag{migdal},\reftag{review}]
effect. The hope was that the dileptons from the
quark-gluon plasma could be
dominant in a narrow window in the invariant
mass distribution, namely between the end of $\pi^0$ Dalitz decay and
the
beginning of the two $\pi$ annihilation threshold.
In this paper we
summarize the results of
this analysis [\reftag{cgr1},\reftag{cgr2},\reftag{cgr3},\reftag{gr}].
Our analysis is mainly concerned with the radiation of photons,
medium effects on the
radiation of gluons have been considered recently in
Ref.[\reftag{wang}].
Our calculations show that the effect is important
for invariant masses smaller than $\rho$-meson mass when the
temperature
is above  150 MeV.
It reduces the dilepton production rate from bremsstrahlung more than
an
order of magnitude. For masses heavier than 1 GeV it
becomes negligible. The origin of
this result is intuitively clear:
radiation being of
electromagnetic origin is relatively slow while the time between
hadronic collisions
becomes very short in a dense medium. It seems natural that the
number of
photons is proportional to the number of collisions but if these
happen
too often the photons emitted at different points of the trajectory
start
to interfere with each other
and the intensity  of radiation is accordingly
 reduced.

For heavy masses the relevant time-scale for the emission of a
lepton-pair is given by its inverse mass,  therefore, heavy mass pairs
are emitted in a very short time interval while light pairs are
emitted
over a much longer time. Thus a dense medium where many
collisions occur, will reduce the bremsstrahlung radiation of low-mass
dilepton pairs but will not affect heavy pairs.

 In [\reftag{cgr1},\reftag{cgr2},\reftag{cgr3}]
the importance
of the  Landau-Pomeranchuk effect
was estimated for virtual photon emission. In [\reftag{gr}] it was
estimated for real photon production.
 The scale of this
effect is determined by the average time between two collisions.
In our calculation we assume that the system is a thermalized static
gas.
We have also used a classical soft photon approximation.
 For low invariant masses and
high temperatures the effect is substantial and will reduce the rate
by a
big factor.

We first present the case where the velocities before and after
scattering
are not correlated, this will be a good approximation in e.g. a pion
gas.
We will then consider the case where the velocities are highly
correlated,
this will be relevant for the case where scattering is dominated by
small
angle forward scattering.
\section{Landau-Pomeranchuk Effect}

\subsection{Quantitative Estimate}
The starting point is the following textbook equation for the energy
radiated per unit of momentum
[\reftag{jackson}] :
$$
{dI\over d^3k}={\alpha\over (2\pi)^2}\left| \int_{-\infty}^{\infty}
 dt e^{i(\omega t -\vec{k}\cdot\vec{r}(t))}
\vec{n}\times \vec{v} \right| ^2
\eqno(1)
$$
where $\vec{r}(t)$ describes the trajectory of the charged particle
and
$\vec{v}(t)$ is its velocity and
$$
\vec{n}= \vec{k} /\omega
\eqno(2)
$$
If only one collision occurs and if the velocity is a constant before
($v_1$) and after the collision ($v_2$)
the energy radiated  is given by
$$
\eqalignno{
{dI\over d^3k}&={\alpha \over (2\pi)^2}\left| \int_{-\infty}^{0}
        dt e^{i(\omega -\vec{k}\cdot\vec{v}_1)t}
\vec{n}\times \vec{v}_1
      + \int_{0}^{\infty}
      dt e^{i(\omega -\vec{k}\cdot\vec{v}_2)t }
   \vec{n}\times \vec{v}_2 \right| ^2&\cr
    &={\alpha \over (2\pi)^2}\left| {\vec{n}\times\vec{v}_1\over
   \omega-\vec{v}_1\cdot\vec{k} }
-{\vec{n}\times\vec{v}_2\over
\omega-\vec{v}_2\cdot\vec{k} }\right|^2&(3)\cr
}
$$
where one recognizes the two propagators familiar from quantum
electrodynamics: $ (p+k)^2-m^2 = 2p_0k_0-2\vec{p}\cdot\vec{k} =
2p_0[\omega - \vec{v}\cdot\vec{k}]$.

It is well-known that the classical expression coincides with the
result
from quantum field theory in the soft photon limit (see e.g. Ref.
[\reftag{pisut}]).

If however a series of collisions occurs,
one has to change the above to a sum over all pieces of
the trajectory
$$
\eqalignno{
{dI\over d^3k}&={\alpha \over (2\pi)^2} \biggl| \int_{-\infty}^{t_1}
 dt e^{i(\omega t -\vec{k}\cdot\vec{r}(t))} \vec{n}\times \vec{v}_1
&\cr
&+ \int_{t_1}^{t_2}
 dt e^{i(\omega t-\vec{k}\cdot\vec{r}(t))} \vec{n}\times \vec{v}_2
&\cr
&+ \int_{t_2}^{t_3}
 dt e^{i(\omega t-\vec{k}\cdot\vec{r}(t))} \vec{n}\times \vec{v}_3
&\cr
&+\cdots & \cr
&+ \int_{t_{n-1}}^{t_n}
 dt e^{i(\omega t-\vec{k}\cdot\vec{r}(t))} \vec{n}\times \vec{v}_n +
\cdots  \biggr| ^2 &(4)\cr
}
$$
Assuming the velocity is constant between two collisions, this leads
to
$$
{dI\over d^3k}={\alpha \over (2\pi)^2}
\left| \sum_{j=1}^N
{\vec{n}\times\vec{v}_j\over \omega -\vec{k}\cdot\vec{v}_j}
 e^{i(\omega t_{j-1}-\vec{k}\cdot\vec{r}_{j-1})}
\left( 1 - e^{-i(\omega -\vec{k}\cdot\vec{v}_j)(t_j-t_{j-1})}\right)
\right|^2
\eqno(5)
$$
where $N$ is the total number of collisions.

A remark :
instead of looking at the radiated energy at infinity we look at the
produced lepton pairs inside the
plasma or  gas, one does not include the first  nor the last leg
in the above sum since they come from contributions radiated at
infinity.
The square leads to
$$
\eqalignno
{
{dI\over d^3k} &=  {\alpha \over (2\pi)^2} \sum_{j=1}^N
    {v_j^2-(\vec{n}\cdot\vec{v}_j)^2\over
    (\omega-\vec{k}\cdot\vec{v}_j)^2}                           &\cr
&  \left[ 2-e^{i\xi_j(\omega-\vec{k}\cdot\vec{v}_j)}
   -e^{-i\xi_j(\omega -\vec{k}\cdot\vec{v}_j)} \right]
&\cr
&  +2{\rm Re}\sum_{j>l}^N
  {\vec{v}_j\cdot\vec{v}_l-
(\vec{n}\cdot\vec{v}_j)(\vec{n}\cdot\vec{v}_l)
   \over (\omega -\vec{k}\cdot\vec{v}_j)(\omega -
\vec{k}\cdot\vec{v}_l)}
   \exp\left[i\sum_{i=l}^{j-1}\xi_i(\omega -\vec{k}\cdot\vec{v}_i)
  \right] &\cr
&  \left( 1 - e^{-i(\omega -\vec{k}\cdot\vec{v}_j)\xi_j}\right)
   \left( 1 - e^{ i(\omega -\vec{k}\cdot\vec{v}_l)\xi_l}\right)
&(6)\cr
}
$$
where
$$
\xi_j = t_j-t_{j-1}
\eqno(7)
$$
As the hadronic gas is taken to be in thermal equilibrium, many
possible
velocities can result in between two collisions. For this reason
the non-diagonal terms in the above equation will give zero
contribution
after one averages over all velocities, provided of
course that no correlations exist between the velocity before and
after
the collision. Only the diagonal terms remain in this case.
We consider this to be applicable to the pion gas case.

The time betwen two successive collisions is given by $\xi$. Taking an
average over the time  between two collisions
can be obtained using the following
distribution

$$
{dW\over d\xi} = a e^{-\xi a}
\eqno(8)
$$
where $a$ is the average time between two collisions and has
been defined in Eq. (1). This is clearly the most natural
distribution.

The average energy radiated away  is given by
$$
{dI\over d^3k}
=N{\alpha \over (2\pi)^2} \left< a\int_0^\infty d\xi e^{-\xi a}
  {v^2-(\vec{n}\cdot\vec{v})^2\over
    (\omega-\vec{k}\cdot\vec{v})^2}
  \left[ 2-e^{i\xi(\omega-\vec{k}\cdot\vec{v})}
   -e^{-i\xi(\omega -\vec{k}\cdot\vec{v})} \right]\right>.
\eqno(9)
$$
The averaging refers to the velocities $\vec{v}$.
The integral over the time between two collisiions, $\xi$, is done
using
$$
a\int_0^\infty d\xi e^{-\xi a}[1-\cos\xi b] =
{b^2\over a^2 +b^2}.
\eqno(10)
$$
This replaces the propagator factor $(\omega-\vec{v}\cdot\vec{k})^2$
by a
regularized
expression

\noindent $(\omega-\vec{v}\cdot\vec{k})^2+a^2$, thus the
divergence due to soft photons is regularized by the mean free time
between two collisions $a$.

To relate the above expressions to cross-sections, we note the
following
$$
\eqalignno{
{dI\over d^3k}&=\omega{dN^{\gamma^*}\over d^3k}& \cr
              &=N\omega{1\over\sigma_{\pi
\pi}}{d\sigma^{\gamma^*}\over
d^3k}.&(11) \cr
}
$$
We are then left with
$$
\left< {d\sigma^{\gamma^*}\over d^3k}\right>  = \sigma_{\pi \pi}
{1\over\omega}
{2\alpha \over (2\pi)^2}
\left< v^2
{(1-\cos^2\theta)\over a^2 + (\omega -kv\cos\theta)^2}\right>
\eqno(12)
$$
the averaging is understood to be over the
velocities between successive
collisions.
For the angular integral we use
$$
\int_{-1}^{+1} dY {1-Y^2 \over a^2 + (\omega - kvY)^2}
= {1\over k^3v^3}\phi
\eqno(13)
$$
where the function $\phi$ is given by
$$
\eqalignno{
\phi &=
 {a^2+k^2v^2-\omega^2\over a} \biggl(
\arctan {\omega +kv\over a} - \arctan {\omega -kv \over a} \biggr)
&\cr
&+ \omega \ln {a^2+(\omega +kv)^2\over a^2+(\omega -kv)^2}
-2kv. &(14)\cr
}
$$
\subsection{Time between two collisions}
The most important
parameter characterizing the Landau-Pomeranchuk effect is the
average time between two collisions. We study here the
inverse of this quantity and denote it by $a$, it can be calculated
from
$$
a\equiv n\sigma v
\eqno(15)
$$
where $n$ is the density of the medium which depends on its
composition and  temperature $T$. For simplicity we will
consider here only a pion gas. In eq. (1)
$\sigma$ denotes the
cross-section of a test pion having velocity $v$
interacting with the particles of
the medium. Thus $a$ is  a function of $T$ and of the momentum of the
test particle. We do not include any dependence on chemical
potentials
as this
 is  inessential for our  considerations.

To calculate $a$ we consider the following expression
$$
a(p,T)=g\int {d^3q\over (2\pi )^3} e^{-E/T}\sigma_{\pi\pi}(s)v,
\eqno(16)
$$
where the relative velocity $v$ is given by
$$
v\equiv |\vec{v}_{12}| = \sqrt{s(s-4m_{\pi}^2)}/2E_1E_2,
\eqno(17)
$$
with $s\equiv (p+q)^2$ and $g$
is the degeneracy factor (3 for the pion gas under consideration).
For the cross-section $\sigma_{\pi\pi}$
we consider  different contributions [\reftag{crs}]:
\item{1)}the low-energy part determined by chiral symmetry :
$$
\sigma_{\pi\pi} (s)={2\over 3}{1\over F_{\pi}^4}{s\over 16\pi^2}
\left[ 1-{5m_\pi^2\over s}+{7m_{\pi}^4\over s^2} \right],
\eqno(18a)
$$
with the pion decay constant $F_{\pi}=0.098$ GeV
\item{2)}the $\rho$-pole contribution
$$
\sigma_{\pi\pi} (s) = {g_{\rho\pi\pi}^4\over 48\pi s}
{ (s-4m_{\pi}^2)^2 \over (s-m_{\rho}^2)^2 +\Gamma_{\rho}^2m_{\rho}^2},
\eqno(18b)
$$

The asymptotic behaviour is taken to be constant, its precise
value is not very important for the range of temperatures we are
considering. It was chosen to be $\sigma_{\pi\pi} (s) \simeq 15{\rm
mb }$.

The resulting values for $a^{-1}$ are shown in figure 2 as a function
of
momentum for different values of the temperature.\par
\medskip
%
%
\indent Figure 2 : Inverse
value of $a$ for different values of the temperature as a function
of the momentum of the incoming particle.
$a$ is the average time between two collisions : $a\equiv n\sigma v$.
\bigskip
As one can see the values for $a^{-1}$ decrease rapidly for high
temperatures.
For a large range of values of  incoming momenta $a$ remains
approximately constant. Below
temperatures of about 150 MeV the values of $a^{-1}$ are too large to
be
consistent with a hadronic gas interpretation, e.g. the mean free
path for
pions having $T\simeq 100$ MeV is of the order of 10 $fm$, this means
that
we cannot expect thermalization of pions.

As we will show below the value of $a$ is essential for the
importance of the
Landau-Pomeranchuk effect: if $a$ is zero, the effect is absent.

In the subsequent calculations we will focus mainly on
comparing results for $a=0$ and for $a$ different from zero, i.e.
respectively without and with the Landau Pomeranchuk effect.
\subsection{Dilepton Rate in a Pion Gas}
We can now relate the expression derived previously
to dilepton pair production  using the standard expression
$$
{d\sigma^{+-}\over dM^2}={\alpha\over 3\pi}{1\over M^2}
\sqrt{1-{4m_l^2\over M^2}}\left( 1+{2m_l^2\over M^2}
\right)\sigma^{\gamma^*}
\eqno(19)
$$
For the order of magnitude estimates we ignore the various threshold
factors in the following, this is certainly justified for dielectron
production.
The expression for the rate is given by kinetic theory as being
(we assume Boltzmann approximation)
$$
{dN^{+-}\over dM^2d^4x}=\int d^3p_1\int
d^3p_2~f(\vec{p}_1)f(\vec{p}_2)
|\vec{v}_{12}|{d\sigma^{+-}\over dM^2}
\eqno(20)
$$
using Eq. (17) for the relative velocity this becomes
$$
\eqalignno{
{dN^{+-}\over dM^2d^4x}=&{d\over (2\pi )^4}{\alpha^2\over 3
\pi^2}{1\over M^2}
\int_{m_{\pi}}^{\infty}p_1dE_1\int_{m_{\pi}}^{\infty}p_2dE_2
 e^{-(E_1+E_2)/T} &\cr
&\int_{-1}^{+1}d(\cos\theta )
\sqrt{s(s-4m_{\pi}^2)}{1\over v_1}\sigma_{\pi\pi}(s)\int_M^{\Delta}
{d\omega\over k^2}\phi&(21)\cr
}
$$
where $d$ is a corresponding degeneracy factor.
We  now discuss, quantitatively, the influence of the
Landau-Pomeranchuk effect on the bremsstrahlung of dilepton pairs in
a hot
hadronic gas. We shall assume for simplicity that the system is
composed of
only pions. The temperature and other thermodynamic quantities are
taken
as time independent.
In figure (3)  we show the dilepton production rate as a function
of the invariant mass of the pair at a temperature of 200 MeV.
It is seen from the figure that, in the low mass region, the
Landau-Pomeranchuk effect decreases the bremsstrahlung
rate by up to two full orders of magnitude for the temperature of 300
MeV.
\par
\epsfxsize=10truecm
\epsffile{berk3.ps}
Figure 3 : Rate as a function of the invariant mass of the dilepton
pair at
a fixed temperature ($T = 200$ MeV) with and without the
Landau-Pomeranchuk effect.
\bigskip
The change of  slope observed in figure 3  at $M\simeq 0.4$ GeV
is related to the
peak structure of the pion-pion elastic cross-section coming
from the $\rho$-meson.

{}From the above properties one can conclude that for  low-mass dilepton
production in relativistc heavy-ion collisions one cannot neglect the
Lan\-dau-Po\-me\-ran\-chuk effect as it leads to an important
reduction of the
bremsstrahlung contribution.
\section{Landau-Pomeranchuk Effect in a Quark-Gluon Plasma}
\par
In order to discuss the role of the Landau-Pomeranchuk effect on
photon
production  due to quark bremsstrahlung in a QGP we follow
the semi- classical approach developed in Ref. [\reftag{migdal}].
\par
Since the successive velocities are strongly correlated we rewrite the
starting equation (1) as follows
$$
\eqalignno{
{dI\over d^3k}&=
 {e_q^2 \alpha\over (2\pi)^2} 2{\rm Re} \int_{0}^{\infty}dt
\int_{0}^{\infty}d\tau
e^{-i[\omega \tau -\vec{k}\cdot(\vec{r}(t+\tau)-\vec{r}(t))]}&\cr
&[\vec{v}(t+\tau)\cdot\vec{v}(t)
-(\vec{n} \cdot \vec{v}(t+\tau))(\vec{n}\cdot\vec{v}(t))]
&(22)\cr
}
$$
assuming that the particle starts to move at the initial time
$t_0=0$.
Eq. (22) describes the energy emitted from a given particle trajectory
$\vec{r}(t)$ . The above form is more suitable for the case where
Coulomb-like small angle  scattering processes dominate. We assume
this to
be the case in a quark-gluon plasma. In the pion gas we assumed that
there
was no correlattion between the changes in velocities.
\par
It is clear from eq. (22), that  one obtains a  non-vanishing result
 even when the test particle
is moving in the vacuum.
This is because  the
radiation   from an
 infinitely  accelerated
particle at the initial time $t_0$ = 0 is included.
Since in our further discussion we shall  only be
 interested  in  medium effects on photon
radiation,  the above vacuum contribution will be subtracted
from our final result.
One has to note also, that, as the time integration in eq. (22) is
carried
out from zero  to infinity, we  explicitly assume that the lifetime
and
the volume of the plasma are infinite.
It will be shown, that
the multiple scattering of a charged particle in
 a medium decreases the probability of  soft, real
or virtual, photon emission. Taking into account the finiteness of
the plasma
volume one would get an additional
 suppression for photon radiation.
In the present  discussion, however, we
 shall not include  the influence of the plasma size
on the photon production rates. Thus, our results should be only
considered as a lower bound for the suppression of dilepton
and photon production in a QGP due to the medium.
\par
Following  Ref. [\reftag{migdal}], one obtains
after subtracting the vacuum
 contribution  finally the following
result for
the
number of virtual photons emitted per unit frequency interval
and per unit time by a fast
charged particle moving in a medium [\reftag{cgr2}]
$$
{dN^{\gamma^* } \over d\omega dt} =
{e^2_q\alpha k \over  \pi \omega}({m_q^2 \over p_1^2}
+{M^2 \over \omega^2}) \int^{\infty}_0 dx  \Bigl({1 \over {\rm
tanh}x} - {1 \over x}\Bigr) e^{-2sx} \sin 2sx
\eqno(23)
$$
where
$$
s = {1 \over 8} \Bigl({m_q^2 \over p_1^2}
+{M^2 \over \omega^2}\Bigr) \sqrt{4\omega \over <\theta_S^2>}
\eqno(24)
$$
and
 $<\theta_S^2>$ is the mean square
of scattering angle of the fast particle per unit path:
$$
<\theta_S^2> \equiv \left< n v_{12} \int d\Omega \theta^2 {d\sigma
\over d\Omega} \right>
\eqno(25)
$$
and $n$ is the density of the medium, $v_{12}$ is the relative
velocity, ${d\sigma \over d\Omega}$ is the elastic cross-section and
$\theta$ is the angle between the initial and the final particle
momentum.
The average $<\theta_S^2>$ depends on both  particle's momentum and
the medium properties.

In the limit when the  virtual photon
 is going on shell, $M \rightarrow 0$, eq.(23) coincides with the
result
 obtained previously by
Migdal [\reftag{migdal}] for the real photon production.
 It is interesting to note, as we are showing later, that the
rates in eq.(23) is  infrared stable in the limit of massless quark.
This is because the invariant mass of a virtual photon acts as an
infrared cut-off in the above equation.

Following Ref. [\reftag{migdal}], let
 us introduce  a new function $\phi (s)$ defined as:
$$
\phi (s) = 24 s^2 \int^{\infty}_0 dx ({1 \over {\rm
tanh}x} - {1 \over x}) e^{-2sx} \sin 2sx
\eqno(26)
$$
which determines the virtual photon distribution in eq.(23).\par
We have found that  the above function can be  approximated by
$$
\phi(s)^* \sim \Bigl[1+\bigl({1 \over 6s}\bigr)^2\Bigr]^{-1/2}
\eqno(27)
$$
The results for the exact eq. (26) and  the approximate eq. (27)
values
of the function $\phi(s)$ are
 within 10-percent
 accuracy  of each other.
\par
The result for the virtual photon emission rate
can finally be written in a much simplified form using Eq. (27) :
$$
{dN^{\gamma^*} \over d\omega dt} =
{2 e^2_q\alpha k \over 3 \pi \omega^2}
<\theta_S^2> \left[ \bigl({m_q^2 \over p_1^2}
+{M^2 \over \omega^2}\bigr)^2
+{4<\theta_S^2> \over 9\omega} \right]^{-1/2}
\eqno(28)
$$
It is easy to see that the above expression remains finite in
the limit of massless quarks. Thus, the scattering of
particles in a medium regularizes the infrared behaviour of the
corresponding amplitudes.
\par
The influence of the
Landau-Pomeranchuk effect on the virtual photon
emission by moving quark in a medium is contained in the
 last term in the square bracket in eq. (28).
If this term  is negligibly small as compared
to the first one, then the effect is absent.
In figure (4) we show the effect quantitatively.\par
\par
Figure 4 : Dielectron
production rate in a quark-gluon plasma with fixed temperature
$T = 0.3$ GeV :
\item{1.)} dot-dashed line: virtual-quark brems\-strah\-lung without
medium effect;
\item{2.)} dashed line: virtual quark brems\-strah\-lung with the
Landau-Pomeranchuk effect taken explicitly into account;
\bigskip
\section{Hydrodynamic Expansion}
\medskip
We have also analyzed the behaviour of the low mass dilepton
spectrum in central heavy ion collisions at the LHC energy,
by taking into account
the expansion dynamics of a hot hadronic matter assuming  Bjorken's
hydrodynamical model [\reftag{cgr3}]. The extension of our
calculations
to the model including transverse  flow
would decrease the contributions of the pion phase to the
overall dilepton spectrum.

 A detailed study of virtual
photon bremsstrahlung was performed including   the
Landau-Pomeranchuk suppression in a hot quark-gluon plasma
and in a pion gas. Our results for the thermal spectrum were then
compared
with  Dalitz background.

We have shown that for high initial temperature $T\sim 0.5-0.9$ GeV
the quark-gluon plasma signal could be  discriminated
measuring transversal momentum spectra of low invariant mass
dielectron
pairs (see figure (5)). This kinematic region is however outside the
range of v
validity of our calculations.
For lower temperature
one would need to subtract
 Dalitz $\eta$ decay contribution in order to measure
dilepton production in a quark-gluon plasma.
\bigskip
\noindent
Figure 5 : Dilepton
rates $dN^{e^+e^-}/dydMdp_t^2$ at fixed value of the mass $M=0.15$ GeV
assuming the  initial temperature $T_0=0.9$ GeV and thermalisation
time
$\tau_0=0.07$ fm.
\bigskip
\section{Conclusions}
\medskip
In conclusion, we have shown that the Landau-Pomeranchuk effect
is important
for low mass dilepton pairs. It quickly becomes negligible above
the mass of the $\rho$-meson. It will therefore not be relevant for
discussions of thermal lepton production in the region between the
$\rho$ and $J/\psi$ peaks. For very low mass dileptons it
reduces the production rate by about an order of magnitude.
The Landau-Pomeranchuk does not reduce the background due to
bremsstrahlung from pions sufficiently to make the quark-gluon plasma
contribution appear clearly. We therefore have to conclude that low
mass dileptons will not give any clear signature for the production
of a
quark-gluon plasma.
\noindent
\bigskip\centerline{\sectnfont References}\bigskip
\item{\reftag{vesa})}
          For a recent a review see e.g.: P.V. Ruuskanen, Proc. Quark
Matter '91,
          Gatlinburg, V. Plasil et al. (eds.); H. Satz, Proc. Int.
          Lepton-Photon Symp., Geneva, Switzerland (1991).
\smallskip
\item{\reftag{crs})}
          J. Cleymans, K. Redlich and H. Satz, Z. f. Physik {\bf C52}
          (1991) 517.
\smallskip
\item{\reftag{landau})}
          L. Landau and I. Pomeranchuk,
          Doklady Akad. Nauk S.S.S.R.{\bf 92}, No. 3, 535; No.4, 735
(1953)
\smallskip
\item{\reftag{migdal})}
          A. B. Migdal,
          Doklady Akad. Nauk S.S.S.R.{\bf 96}, No. 1, 49 (1954);
          Phys. Rev. {\bf 103}, 1811 (1956).
\smallskip
\item{\reftag{review})}
      For a review see e.g. :
      A.I. Akhiezer and N.F. Shulga, Sov. Phys. Usp. {\bf 30}
          197 (1987)
\smallskip
\item{\reftag{cgr1})}
          J. Cleymans, V.V. Goloviznin and K. Redlich,
          Phys. Rev. {\bf D47} (1993) 173.
\smallskip
\item{\reftag{cgr2})}
          J. Cleymans, V.V. Goloviznin and K. Redlich,
          Phys. Rev. {\bf D47} (1993) 989-997.
\smallskip
\item{\reftag{cgr3})}
          J. Cleymans, V.V. Goloviznin and K. Redlich,
          Z. f. Physik {\bf C} (to be published, 1993).
\smallskip
\item{\reftag{gr})}
          V.V. Goloviznin and K. Redlich,
          Z. f. Physik {\bf C} (to be published, 1993).
\smallskip
\item{\reftag{wang})}
          X.N. Wang and M. Gyulassy,
          Phys. Rev. {\bf D} (1993).
\smallskip
\item{\reftag{jackson})}
          {\sl Classical Electrodynamics, }
          J. D. Jackson, equation (14.67)
          J. Wiley \& Sons, Inc., New York.
\smallskip
\vfil\eject\end